\documentclass[aps,prl,superscriptaddress,twocolumn,nopacs,amsmath,amssymb,letter]{revtex4-1}

\usepackage{color}
\usepackage{graphicx}% Include figure files
\usepackage{dcolumn}% Align table columns on decimal point
\usepackage{bm}% bold math
\usepackage{xspace}

\setcitestyle{numbers,square}

\begin{document}

\title{Itinerant ferromagnetism of the Pd-terminated polar surface of PdCoO$_2$}

\author{F.~Mazzola}
\affiliation {SUPA, School of Physics and Astronomy, University of St. Andrews, St. Andrews KY16 9SS, United Kingdom}

\author{V.~Sunko}
\affiliation {SUPA, School of Physics and Astronomy, University of St. Andrews, St. Andrews KY16 9SS, United Kingdom}
\affiliation {Max Planck Institute for Chemical Physics of Solids, N{\"o}thnitzer Stra{\ss}e 40, 01187 Dresden, Germany}

\author{S.~Khim}
\author{H.~Rosner}
\author{P.~Kushwaha}
\affiliation {Max Planck Institute for Chemical Physics of Solids, N{\"o}thnitzer Stra{\ss}e 40, 01187 Dresden, Germany}

\author{O.~J.~Clark}
\author{L.~Bawden}
\affiliation {SUPA, School of Physics and Astronomy, University of St. Andrews, St. Andrews KY16 9SS, United Kingdom}

\author{I.~Markovi{\'c}}
\affiliation {SUPA, School of Physics and Astronomy, University of St. Andrews, St. Andrews KY16 9SS, United Kingdom}
\affiliation {Max Planck Institute for Chemical Physics of Solids, N{\"o}thnitzer Stra{\ss}e 40, 01187 Dresden, Germany}

\author{T.~K.~Kim}
\author{M.~Hoesch}
\affiliation{Diamond Light Source, Harwell Campus, Didcot, OX11 0DE, United Kingdom}

\author{A.~P.~Mackenzie}
\email{andy.mackenzie@cpfs.mpg.de}
\affiliation {Max Planck Institute for Chemical Physics of Solids, N{\"o}thnitzer Stra{\ss}e 40, 01187 Dresden, Germany}
\affiliation {SUPA, School of Physics and Astronomy, University of St. Andrews, St. Andrews KY16 9SS, United Kingdom}

\author{P.~D.~C.~King}
\email{philip.king@st-andrews.ac.uk}
\affiliation {SUPA, School of Physics and Astronomy, University of St. Andrews, St. Andrews KY16 9SS, United Kingdom}

\begin{abstract}
We study the electronic structure of the Pd-terminated surface of the non-magnetic delafossite oxide metal PdCoO$_2$. Combining angle-resolved photoemission spectroscopy and density-functional theory, we show how an electronic reconstruction driven by surface polarity mediates a Stoner-like magnetic instability towards itinerant surface ferromagnetism. Our results reveal how this leads to a rich multi-band surface electronic structure, and provide spectroscopic evidence for an intriguing sample-dependent coupling of the surface electrons to a bosonic mode which we attribute to electron-magnon interactions. Moreover, we find similar surface state dispersions in PdCrO$_2$, suggesting surface ferromagnetism persists in this sister compound despite its bulk antiferromagnetic order.
\end{abstract}

\date{\today}

\maketitle

The ability to modulate the collective properties of correlated electron systems at their interfaces and surfaces underpins the burgeoning field of ``designer'' quantum materials~\cite{mannhart_oxide_2010,zubko_interface_2011,hwang_emergent_2012,gozar_high-temperature_2008,chakhalian_magnetism_2006,reyren_superconducting_2007,boris_dimensionality_2011,monkman_quantum_2012,king_atomic-scale_2014,wang_tailoring_2016}. 
Most attention to date has been focused on perovskite-based transition-metal oxides~\cite{mannhart_oxide_2010,zubko_interface_2011,hwang_emergent_2012}, but it is important to expand the search of materials systems which may be tuned to host new surface and interface phases.  Here, we focus on the ``ABO$_2$'' delafossite oxides~\cite{shannon_chemistry_1971,shannon_chemistry_1971-1,shannon_chemistry_1971-2,tanaka_growth_1996,takatsu_roles_2007,mackenzie_properties_2017}, a particularly promising material class both because of its naturally layered structure, as well as the potential to drastically alter its physical properties by changing the A- and B- site cations~\cite{Cheong:2007aa,kawazoe_p-type_1997,Singh_Electronicandthermoelectric_2007}.
MCoO$_2$ [M=Pt,Pd] are non-magnetic metals with simple single-band Fermi surfaces. In bulk, these stand apart as the most conductive of all known normal-state oxides~\cite{tanaka_growth_1996,hicks_quantum_2012,takatsu_extremely_2013,kushwaha_nearly_2015,mackenzie_properties_2017}. 
The polar nature of their surfaces, however, opens the potential to stabilise local electronic environments and phases different to those of the bulk~\cite{kim_fermi_2009,noh_anisotropic_2009,sunko_maximal_2017}. Using angle-resolved photoemission spectroscopy (ARPES) we show how this drives an intrinsic Stoner instability towards ferromagnetism at the Pd-terminated surface of both PdCoO$_2$, and of its sister compound PdCrO$_2$, a bulk antiferromagnet.

%====================================
\begin{figure}[!b]
\includegraphics[width=\columnwidth]{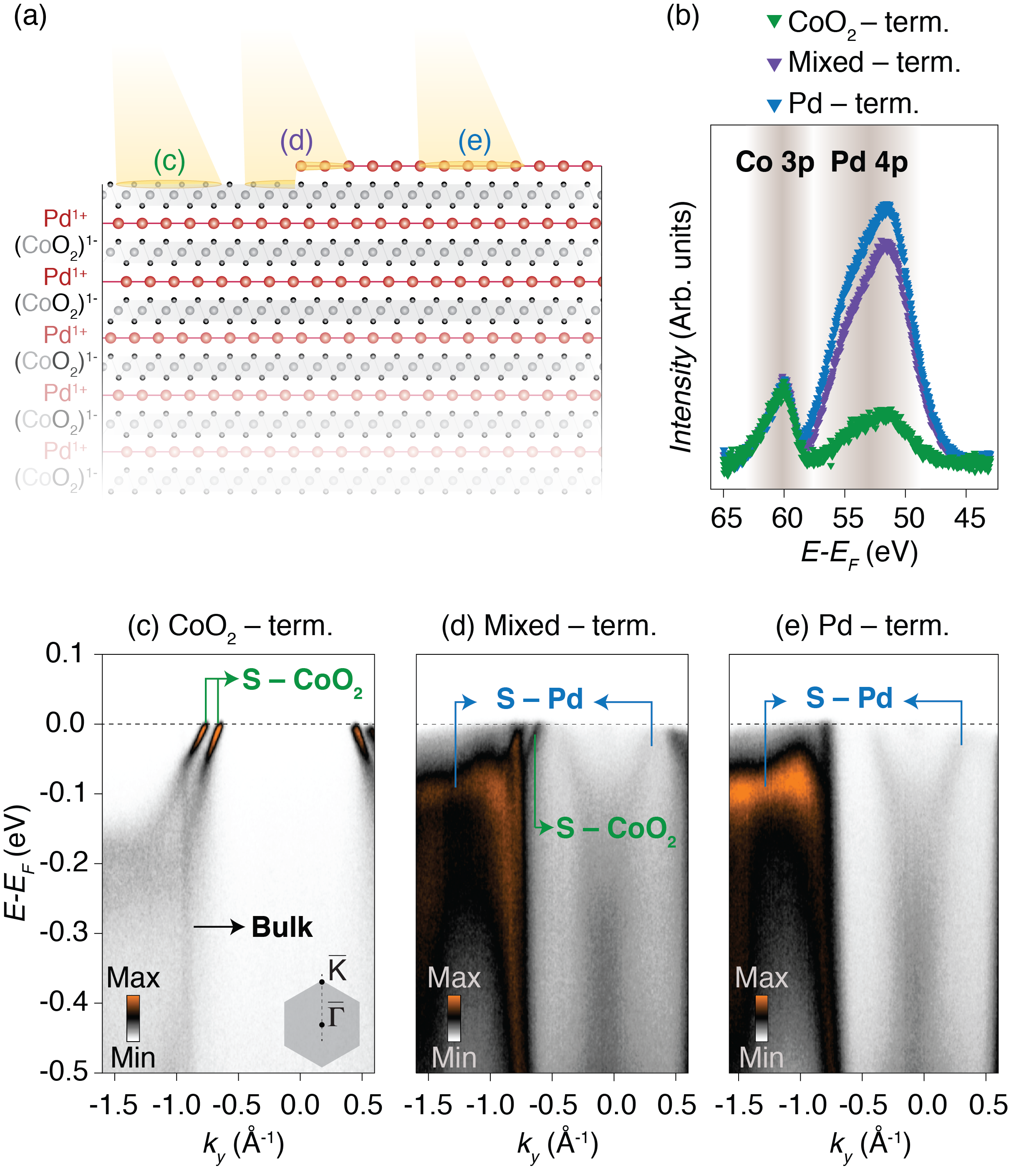}
\caption{(a) Side view of the crystal structure of PdCoO$_2$, showing two possible surface terminations which can be expected. (b) XPS spectra ($h\nu=120$~eV, after subtraction of a linear background and normalised by the area of the Co~3p peak) at different spatial locations of a cleaved crystal show varying relative ratios of Co and Pd core level peaks. (c-e) Markedly different electronic structures are observed by ARPES ($h\nu=90$~eV, \textit{p}-polarisation, measured along $\overline{\Gamma}-\overline{K}$) at these locations, corresponding to predominantly (c) CoO$_2$, (e) Pd and (d) more mixed surface terminations.}
\label{f:termination}
\end{figure}
%=====================================

Single-crystals of PdCoO$_2$ and PdCrO$_2$ were grown by a flux method in sealed quartz tubes \cite{tanaka_growth_1996,Takatsu_single_2010}. These were cleaved {\it in situ} at the measurement temperature of $T\sim\!10$~K. ARPES measurements were performed at the I05 beamline of Diamond Light Source, UK, using a Scienta R4000 electron analyser and variable light polarisations with photon energies between 60 and 120~eV. Density functional theory (DFT) calculations were performed using the full-potential FPLO code~\cite{koepernik_full-potential_1999,opahle_full-potential_1999,fplo}, utilising the Perdew-Burke-Ernzerhof~\cite{perdew_generalized_1996-1} formalism and including spin-orbit coupling. The surface electronic structure was calculated employing a symmetric slab containing 9 Pd layers, with a vacuum gap of 15~\AA. For the inner layers of the slab, the bulk experimental crystal structure~\cite{kushwaha_nearly_2015} was used, while the surface atomic positions were relaxed. Further details of the influence of this surface relaxation, and of applying a $+U$ correction on the Co orbitals, are described in the Supplemental Information (Figs. S1 and S2)~\cite{Supplementary_PdCoO2_Pd_surf}.

PdCoO$_2$ is comprised of triangular-lattice metallic Pd planes separated by insulating CoO$_2$ layers. This crystal structure has natural cleavage planes above and below each Pd layer: two inequivalent surface terminations would therefore be expected (Fig.~\ref{f:termination}(a)). Consistent with this, we find a strong variation in the relative intensity of Co~$3p$ and Pd~$4p$ core levels measured by  X-ray photoelectron spectroscopy (XPS) at different patches of the cleaved sample surface (Fig.~\ref{f:termination}(b)). This is correlated with a marked difference in the electronic structure measured by ARPES (Fig.~\ref{f:termination}(c-e)). Across the sample, we find a steeply-dispersive state which we attribute as the Pd-derived bulk band~\cite{noh_anisotropic_2009}. In Fig.~\ref{f:termination}(c), our measurements additionally show a pair of massive hole-like bands (Fig.~\ref{f:termination}(c), labeled `S-CoO$_2$'). These have been assigned previously as surface states from the CoO$_2$-termination \cite{noh_anisotropic_2009,sunko_maximal_2017}. This is consistent with our XPS measurements from the same sample region, which yield the greatest ratio of Co:Pd core-level spectral weight. We have recently shown how such surface states host a surprisingly-large Rashba-like spin splitting arising due to a large energy scale of inversion symmetry breaking at this surface~\cite{sunko_maximal_2017}. 

Measurements from a different patch of the same sample (Fig.~\ref{f:termination}(e)) reveal a completely different surface electronic structure, which we describe in detail below. Our XPS measurements exhibit a much larger spectral weight of the Pd than the Co-derived core-level peak for this sample region (Fig.~\ref{f:termination}(b)), and we thus attribute these as surface states originating from the Pd-terminated surface (`S-Pd'). We note that for most regions where such states are visible, we observe a superposition of these spectral features with those of the CoO$_2$-terminated surface (e.g., Fig.~\ref{f:termination}(d)). This indicates a rather limited spatial extent of typical Pd-terminated regions, with a mixed surface termination within our probing light spot area ($\sim\!50\;\mu$m diameter). In the following we show our highest-quality ARPES data, obtained from three samples (see also Supplemental Fig.~S4~\cite{Supplementary_PdCoO2_Pd_surf}) which all exhibit such mixed surface terminations.

%===================================
\begin{figure}
\includegraphics[width=\columnwidth]{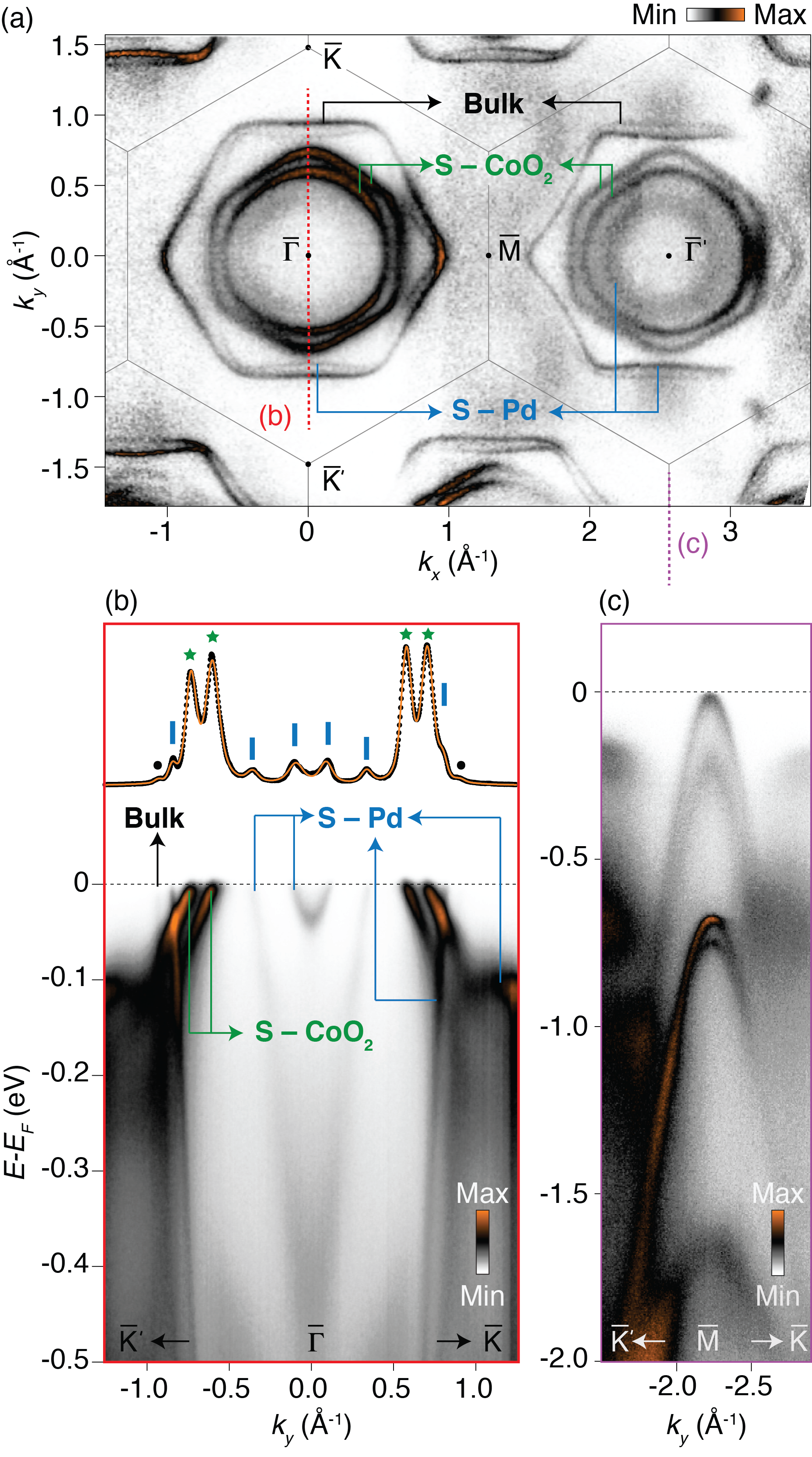}
\caption{(a) ARPES Fermi surface ($h\nu=110$~eV, $E_F\pm{15}$~meV, \textit{p}-pol.) and dispersions ($h\nu=90$~eV, \textit{p}-pol.) measured along (b) $\overline{\Gamma}-\overline{K}$ and (c) $\overline{M}-\overline{K}$. As well as the bulk band (Bulk), a series of additional surface states arising from the Pd-terminated (S-Pd) and CoO$_2$-terminated (S-CoO$_2$) surfaces are visible. A momentum distribution curve (MDC) at the Fermi level ($E_F\pm{5}$~meV) is shown in (b), together with a 12-band peak fit (orange line) and corresponding Fermi momenta (marked by dashes, stars and dots, for S-Pd, S-CoO$_2$ and bulk, respectively) of such bulk and surface states which cross $E_F$.}
\label{f:band_str}
\end{figure}
%=====================================

The measured Fermi surfaces and dispersions are shown in Fig.~\ref{f:band_str}. The CoO$_2$-terminated surface states form hexagonal and nearly circular hole-like Fermi surfaces about the Brillouin zone centre, consistent with previous observations~\cite{noh_anisotropic_2009,sunko_maximal_2017}. We do not consider these further here. Two larger electron-like Fermi surfaces with sharp linewidths are also visible. From comparison with previous experiments~\cite{noh_anisotropic_2009,hicks_quantum_2012}, we attribute the largest of these as the Pd-derived bulk Fermi surface. The other has a similar topography, but slightly smaller average $k_F$, and has greater spectral weight when measured in the second Brillouin zone (Fig.~\ref{f:band_str}(a)). This band is also evident as a steeply-dispersing state in our measured dispersions (Fig.~\ref{f:band_str}(b)), and appears to be an approximate replica of the bulk state, but shifted towards the Fermi level by $\sim\!430$~meV.  

While it is not surprising for a polar charge to change the binding energy of states localised at a polar surface, simple electrostatic arguments~\cite{Note1} suggest that this should manifest as an effective electron, rather than hole, doping of the Pd-terminated surface here~\cite{kim_fermi_2009}. Moreover, rather than a simple rigid shift of the bulk Pd-derived valence band, our measured dispersions (Fig.~\ref{f:band_str}(b,c)) reveal a much richer surface electronic structure. As well as the steep band discussed above, Fig.~\ref{f:band_str}(b) shows a flat-topped band located $\sim\!100$~meV below the Fermi level, and a pair of electron-like bands crossing $E_F$ near the Brillouin zone centre (all labeled S-Pd in Fig.~\ref{f:band_str}(b)).  The outer of these electron pockets can also be seen in our Fermi surface measurements (Fig.~\ref{f:band_str}(a)), while the innermost band is not clearly observed at the photon energy used.  Additionally, multiple fully-occupied bands are found at the Brillouin zone face $\overline{\mbox{M}}$-point (Fig.~\ref{f:band_str}(c)).

%=====================================
\begin{figure}
\includegraphics[width=\columnwidth]{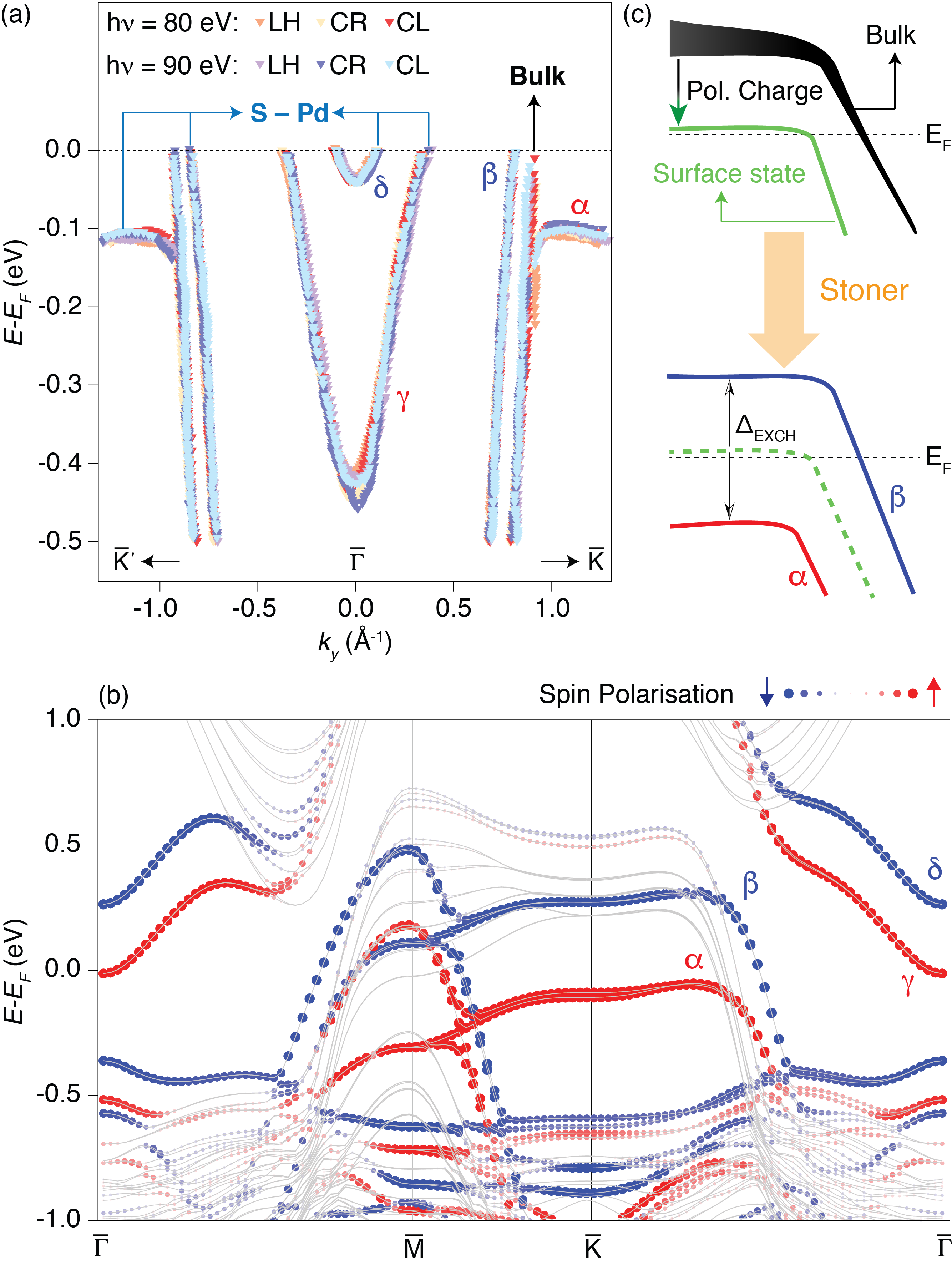}
\caption{(a) Bulk and surface band dispersions exacted from ARPES measurements using different photon energies and light polarisations: linear horizontal (LH, \textit{p}-polarisation), circular right (CR) and circular left (CL) polarisations. (b) Corresponding DFT supercell calculations for a Pd-terminated surface, projected onto majority (red) and minority (blue) spin components of the first PdCoO$_2$ layer, indicating surface ferromagnetism. (c) Schematic illustration of the band structure evolution at the surface, showing electron doping of the surface Pd-layer due to an electronic reconstruction (top). This moves a high density of states to the Fermi level, mediating a Stoner instability and resulting in an exchange splitting of the surface state (bottom).}
\label{f:DFT_comp}
\end{figure}
%=====================================

We show in Fig.~\ref{f:DFT_comp}(a) the dispersions of the bulk and Pd-derived surface states extracted from the measurements shown in Fig.~\ref{f:band_str}(b), as well as additional ones performed under different conditions, confirming the rich multi-band surface electronic structure of PdCoO$_2$. We attribute this to the result of an intrinsic Stoner-like instability to itinerant surface ferromagnetism. Indeed, ferromagnetism has previously been predicted for the Pd-terminated surface of this material~\cite{kim_fermi_2009}, and is also found by our own DFT supercell calculations (Fig.~\ref{f:DFT_comp}(b)) which yield an electronic structure in good qualitative agreement with our measurements. Comparison with the DFT allows us to identify the experimentally observed flat (labeled $\alpha$ in Fig.~\ref{f:DFT_comp}(a,b)) and steep (labeled $\beta$) bands as an exchange-split pair. Similarly, the two electron pockets (labeled $\gamma$ and $\delta$) are also a majority-spin and minority-spin band, respectively. This assignment is consistent with the Luttinger count we extract experimentally: taking the observed bands as spin-polarised, we find $N=0.55\pm0.03$ electrons/unit cell, close to the value of 0.5 that would be expected from the simplest arguments given the polar surface charge. Our measurements thus provide convincing evidence for surface ferromagnetism of the Pd-terminated surface of PdCoO$_2$.

%====================================
\begin{figure*}[!t]
\includegraphics[width=\textwidth]{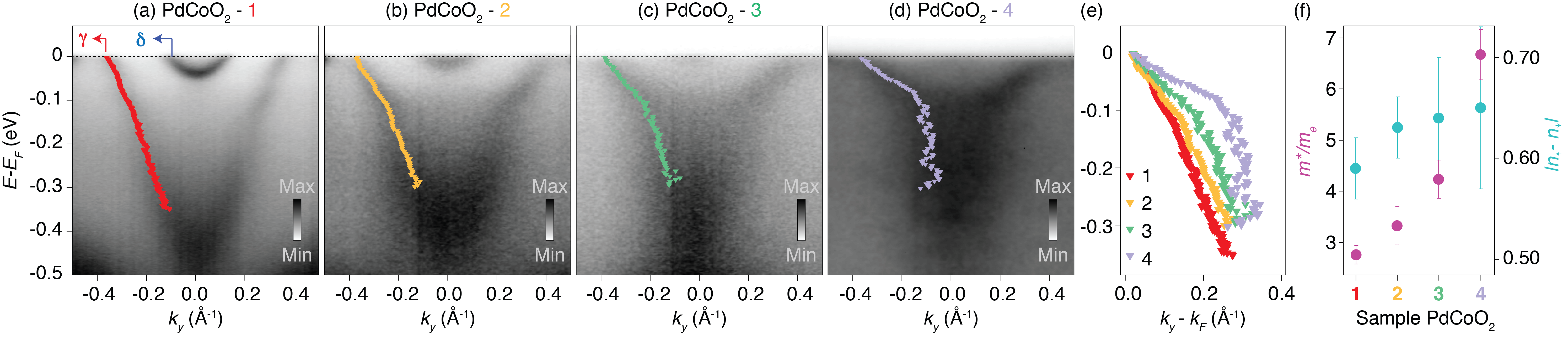}
\caption{(a-d) ARPES measurements of the $\gamma$ and $\delta$ bands from (a,b) different patches of the same sample, and (a,c,d) different samples of PdCoO$_2$ ($h\nu=90$~eV, \textit{p}-pol.). The peak positions extracted from fitting MDCs are shown as coloured markers in (a-d) and are plotted together in (e). (f) Experimental values of the quasiparticle masses extracted from the gradient of these fits close to the Fermi level and the measured Fermi momenta. The difference in the Luttinger area determined for spin-majority and spin-minority bands is also shown in (f).}
\label{f:interactions}
\end{figure*}
%=====================================

The origin of this is shown schematically in Fig.~\ref{f:DFT_comp}(c). The band which forms the bulk Fermi surface, while extremely steep at the Fermi level~\cite{noh_orbital_2009,kushwaha_nearly_2015}, becomes much flatter above $E_F$ (see the flat $k_z$-projected bulk bands along $\overline{\Gamma}-\overline{\mbox{K}}-\overline{\mbox{M}}$ evident in Fig.~\ref{f:DFT_comp}(b)). Compensating the surface polarity should lead to electron doping of the surface layer, as discussed above, creating a surface copy of the bulk band at a higher binding energy as shown in green on the schematic in Fig.~\ref{f:DFT_comp}(c). The high density of states (DOS) associated with its flat band top is therefore pushed towards the Fermi level. The surface DOS thus becomes sufficiently high at $E_F$ to overcome the Stoner criterion, exchange splitting the band into a pair of spin-split $\alpha$ and $\beta$ bands, as observed in both our experiment and calculations (see also Supplemental Fig.~S2 \cite{Supplementary_PdCoO2_Pd_surf}).  Additional near-$E_F$ surface states ($\delta$ and $\gamma$) as well as hole-like bands at $\overline{M}$ also inherit a similar exchange splitting (Fig.~\ref{f:DFT_comp}(b))~\cite{Note2}. We note that the surface magnetism observed here is qualitatively different to the recent observations of ferromagnetism arising due to uncompensated moments at the surface of, for example, antiferromagnetic (AF) EuRh$_2$Si$_2$~\cite{Chikina:2014aa}. Instead, it reflects an intrinsic instability of the underlying electronic structure which can be triggered by a pronounced self-doping of the system in response to its polar surface charge.

From our extracted Luttinger areas, we estimate a total magnetisation, $M=(N^\uparrow-N^\downarrow)\mu_B=(0.59\pm0.03)\mu_B$/unit cell. As the wavefunctions of the surface states are mostly localised on the surface Pd layer (see Supplemental Fig.~S3), this layer also hosts the largest moment (found to be 0.4~$\mu_B$ in our calculations, see also Supplemental Information~\cite{Supplementary_PdCoO2_Pd_surf}). Nonetheless, finite orbital mixing with the subsurface CoO$_2$ block means that the near-surface Co also inherits a finite, but much smaller, moment, found to be 0.1~$\mu_B$ in our calculations. We return to this point below.

As the ferromagnetic state is localised in the topmost layers of the crystal, ARPES is the ideal probe not only to demonstrate its existence, but also to further investigate its many-body properties. In particular, we find that the $\gamma$ band exhibits two apparent ``kinks'' in its measured dispersion (Fig.~\ref{f:interactions}(a) and Supplemental Fig.~S5 \cite{Supplementary_PdCoO2_Pd_surf}), which are characteristic spectroscopic signatures of coupling to bosonic modes~\cite{hufner_very_2007}. The first of these, at an energy of $\sim\!50$~meV, could be attributed to the $E_{g}$ phonon mode of PdCoO$_2$ \cite{Cheng_roleOfPhononPdCoO2, takatsu_roles_2007}. The higher-energy kink, however, is located at $\sim\!200-250$~meV, an energy that is too high for any known phonon modes in PdCoO$_2$~\cite{Cheng_roleOfPhononPdCoO2, takatsu_roles_2007}. Instead, we tentatively attribute this as arising from coupling of the surface state electrons to a magnon, consistent with a self-energy analysis shown in Supplemental Fig.~S5 \cite{Supplementary_PdCoO2_Pd_surf}. 

%====================================
\begin{figure}[!b]
\includegraphics[width=\columnwidth]{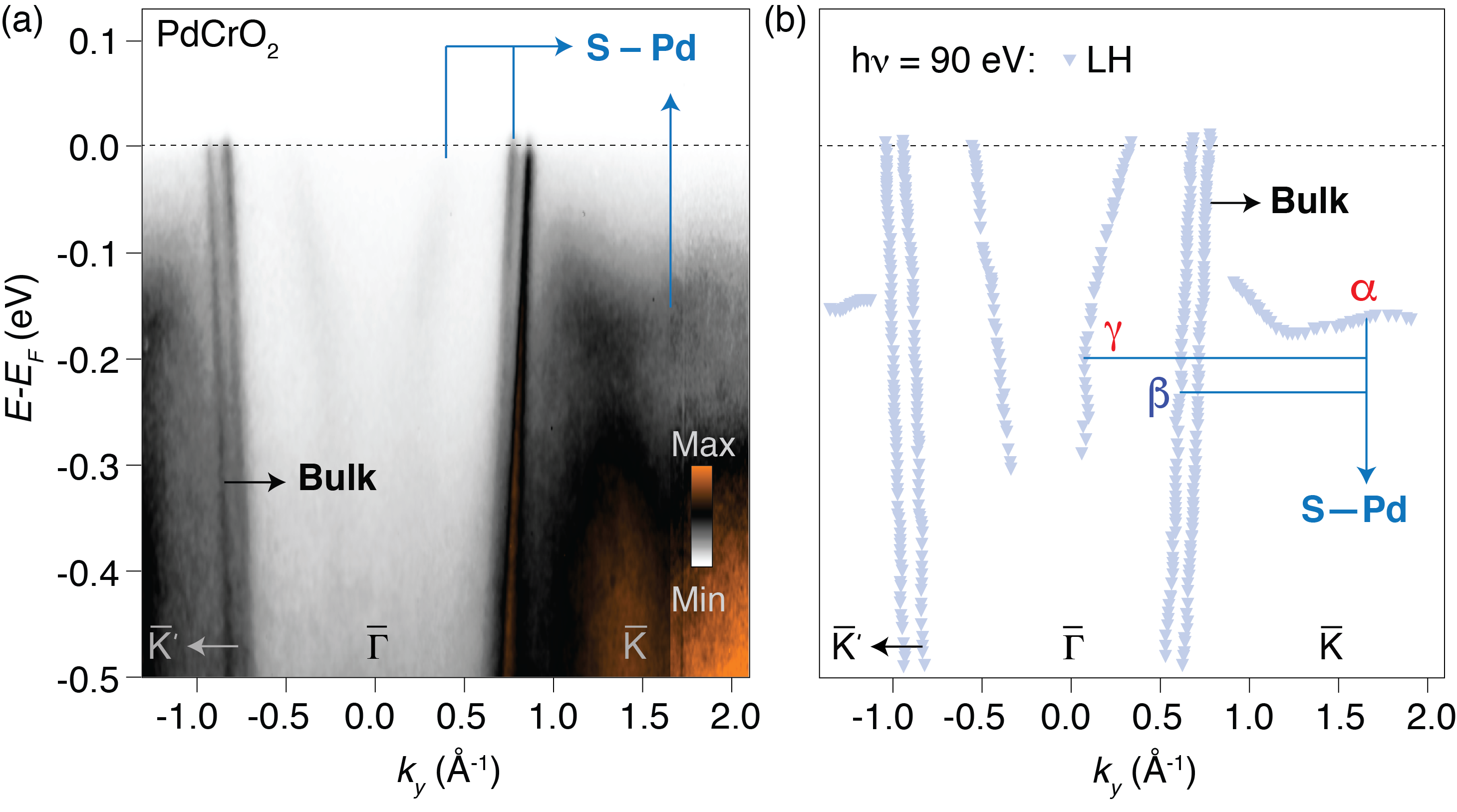}
\caption{(a) ARPES spectra of PdCrO$_2$, measured along the $\overline{\Gamma}-\overline{K}$ direction ($h\nu=90$~eV, \textit{p}-pol.). (b) Corresponding band dispersions extracted from the data shown in (a). The flat $\alpha$ and steep $\beta$ exchange-split pair evident in PdCoO$_2$ (Fig.~\ref{f:DFT_comp}(a)) are also observed here, as well as the electron-pocket of the spin-majority $\gamma$ band. The spin-minority $\delta$ band appears to be fully unoccupied.}
\label{f:PdCrO2}
\end{figure}
%=====================================

Intriguingly, the strength of this kink feature varies substantially across different measured samples (Fig.~\ref{f:interactions}(a,c,d)) and even for different spatial locations of the same sample (Fig.~\ref{f:interactions}(a,b)), causing the quasiparticle mass to grow from $2.8\pm0.2$~$m_e$ to $6.7\pm0.5$~$m_e$ (Fig.~\ref{f:interactions}(e,f)). We cannot rule out that domain size and/or surface disorder change between the different measurements and samples shown here. Experimentally, we find the large mass enhancement is accompanied by a small shift of the surface states towards the Fermi level (Supplemental Fig.~S4(e)~\cite{Supplementary_PdCoO2_Pd_surf}), but negligible change of the total surface carrier density extracted from the measured Luttinger areas, and a barely-resolvable increase in magnetisation (Fig.~\ref{f:interactions}(f)). The origin of these effects remains an interesting question for future study.

Irrespective of this, our work demonstrates how electronic reconstructions at polar surfaces can be exploited to trigger incipient instabilities of the underlying quantum many-body system. Intriguingly, our measurements of the Pd-terminated surface of PdCrO$_2$ (Fig.~\ref{f:PdCrO2}), which hosts local-moment AF order on the Cr sites~\cite{takatsu_critical_2009}, show an extremely similar surface electronic structure as for PdCoO$_2$. This is consistent with previous measurements of the chromate~\cite{sobota_electronic_2013}. Combined with the analysis of PdCoO$_2$ presented here, this strongly suggests that PdCrO$_2$ also supports surface ferromagnetism, naturally forming a ferromagnetic/anti-ferromagnetic heterostructure at its surface. The influence of the resulting magnetic interactions on the subsurface transition-metal layer, which hosts a small ferromagnetic moment in the Co-containing compound but is AF-ordered in the bulk of the Cr-based material, will be an interesting topic for future exploration. 

Targeted engineering of such magnetic competition will be aided by the flexibility of the delafossite oxide series. For example, electron doping on the Pd site of PdCrO$_2$ could be used to drive a Stoner transition of the bulk Pd layers, forming an intrinsic superlattice of two-dimensional itinerant ferromagnets and triangular-lattice local-moment antiferromagnets. Polar interfaces of delafossites with other materials will provide further routes to create and manipulate rich electronic and magnetic phase diagrams in these systems. More generally, our study further highlights the powerful role that polar interfaces can be expected to play in controlling not only the electronic structure, but also the collective phases of designer oxide heterostructures.

\noindent {\it{Acknowledgements:}} We thank C.~Hooley for useful discussions. We gratefully acknowledge support from the European Research Council (through the QUESTDO project), the Engineering and Physical Sciences Research Council, UK (Grant No.~EP/I031014/1), the Royal Society, the Max-Planck Society and the International Max-Planck Partnership for Measurement and Observation at the Quantum Limit. VS, LB, OJC and JMR acknowledge EPSRC for PhD studentship support through grant Nos.~EP/L015110/1, EP/G03673X/1, EP/K503162/1, and EP/L505079/1. We thank Diamond Light Source for access to Beamline I05 (Proposal Nos.~SI12469, SI14927, and SI16262) that contributed to the results presented here.
\linespread{0.9}

\end{document}